# Community-Led AI Integration for Wildfire Risk Assessment: A Participatory AI Literacy and Explainability Integration (PALEI) Framework in Los Angeles, CA

Sanaz Sadat Hosseini[1], Mona Azarbayjani[1], Mohammad Pourhomayoun[2], Hamed Tabkhi[1]

[1]The University of North Carolina at Charlotte, Charlotte, NC
[2]California State University, Los Angeles, CA

ABSTRACT: Climate-driven wildfires continue to intensify in severity, particularly in urban regions like Southern California. However, traditional fire risk communication tools often fail to earn public trust due to inaccessible design, non-transparent outputs, and a lack of contextual relevance. These limitations are particularly problematic in high-risk urban communities, where trust in technological tools depends on how clearly and locally relevant the information is presented. Neighborhoods such as Pacific Palisades, Pasadena, and Altadena in Los Angeles, CA face recurrent wildfire threats and exemplify these challenges. This study introduces a community-led approach for integrating AI into wildfire risk assessment using the Participatory AI Literacy and Explainability Integration (PALEI) framework. PALEI emphasizes early literacy building, value alignment, and participatory evaluation before any predictive model is deployed, prioritizing clarity, accessibility, and mutual learning between developers and residents. Early engagement findings indicate strong acceptance of visual, context-specific risk communication, favorable fairness perceptions across neighborhoods, and clear adoption interest, alongside privacy and data-security concerns that shape trust. Participants emphasized localized imagery, accessible explanations, neighborhood-specific mitigation guidance, and transparent communication of uncertainty. The intended outcome is a mobile app co-designed with users and local stakeholders that enables residents to scan visible property features and receive interpretable fire risk scores, along with customized recommendations. By embedding local context into the design, the tool becomes more than a technical interface: it serves as an everyday resource for risk awareness and preparedness. This study argues that user experience must be treated not as an afterthought but as a central element in ethical and effective AI deployment. By establishing a literacy-first, participatory foundation for AI design, the study provides a replicable pathway for applying the PALEI framework to other regions facing growing climate-related hazards.

PAPER SESSION TRACK: Technologies of Place, Policy as Design Catalyst

## INTRODUCTION & PROBLEM CONTEXT

Increasing climate change, urban expansion, and inadequate mitigation efforts make wildfires a growing threat to communities worldwide (Goswami et al. 2024; Kim et al. 2024). In Southern California, neighborhoods such as Pacific Palisades, Pasadena, and Altadena face recurring wildfire threats. Despite this, existing fire risk assessment and communication tools often fail to earn public trust due to being inaccessible, having opaque outputs, and not being contextually relevant. A key component of trust in high-risk urban communities is the ability of tools to correlate clearly with local conditions and present information in such a way that residents will be able to interpret and act on it easily.

Wildfire vulnerability has many dimensions, but one of the most important and underexplored is the rapid spread of fire from one house to another due to exterior conditions and landscaping (Gonzalez-Mathiesen and March 2018; FEMA 2025). Traditionally, assessment approaches are focused on parcel-level structural characteristics, such as building materials or vegetation within a limited radius, but overlook broader transmission pathways across residential landscapes ("Land Use Planning – Wildfire Risk to Communities" 2025; Mann et al. 2014). As a result of dense vegetation, flammable exterior elements, prevailing winds, and limited buffering between homes, fires can cascade from one property to the next, overwhelming firefighting capacity and evacuation planning (Zhou, Lin, and Wang 2025; Hao et al. 2023; Jin et al. 2020; Penney et al. 2024). Current mitigation strategies, like defensible space guidelines, often address risks in isolation rather than acknowledging neighborhood-scale interdependencies (Mann et al. 2014).

Recent wildfires in Altadena, Pacific Palisades, and Pasadena illustrate this spatial complexity, where some homes were totally destroyed while nearby structures remained intact as a result of differences in their vegetation management, setbacks, and buffers. Figure 1 shows how neighborhood-scale conditions have influenced recent wildfire impacts in Pacific Palisades. There is a general lack of clarity among residents regarding what makes their own properties more or less vulnerable and how current risk assessment tools interpret risk. Existing platforms such as FireCast (Conservation International 2023), Wildfire Analyst (Giglio et al. 2018), and the U.S. Forest Service's Wildfire Risk to Communities portal (Mtani and Mbuya 2018) provide rich technical detail but are designed primarily for experts. These tools lack transparent explanation mechanisms, rely on complex modeling interfaces, and rarely incorporate contextual information from residents, such as photographs or neighborhood observations (Radeloff et al. 2023). This gap highlights the need for approaches that combine technical rigor with participatory communication, accessibility, and explainability.

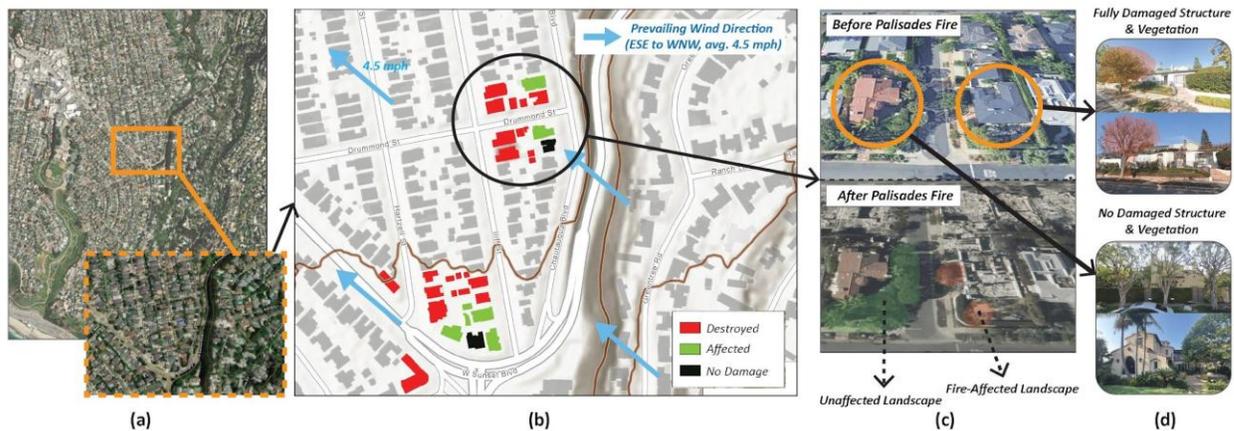

**Figure 1**: Neighborhood-scale wildfire impacts in Pacific Palisades, showing destroyed, affected, and unaffected structures alongside wind direction and pre/post-fire imagery. These patterns illustrate how micro-level conditions contribute to community-level fire spread. Source: (Authors 2025)

To address these limitations and gaps, the main goal of this research is to develop and evaluate a participatory, literacy-centered, and explainable framework for integrating artificial intelligence into wildfire risk assessment. A participatory AI literacy and explainability integration framework (PALEI) is the core methodological contribution of this study. PALEI introduces a human-centered process that embeds AI literacy, fairness, and participatory governance in the early phases of environmental technology design. Rather than emphasizing technical performance alone, PALEI prioritizes clarity, accessibility, and mutual learning among community members to ensure emerging systems remain understandable, contextually grounded, and aligned with community values. The concept of explainability and literacy is treated as fundamental—not optional—elements of AI design, testing, and evaluation.

To evaluate how the PALEI framework functions in a real-world setting, this study applies it to the early design of a wildfire risk assessment tool, focusing first on the priority communities of Altadena, Pasadena, and Pacific Palisades in Los Angeles County—areas that have recently experienced wildfire impacts and face ongoing risk. The research is organized around the initial stages of the PALEI framework, beginning with literacy-building and early engagement activities, including an initial virtual community town hall where residents reviewed conceptual wildfire-risk information and responded to preliminary explanation formats. The event was designed to surface which explanation styles feel intuitive or confusing, how uncertainty should be communicated, and what conditions—privacy, fairness, and neighborhood context—shape residents' trust in emerging AI tools. These early insights will guide the next phases of the PALEI process, including explanation-persona development, co-design workshops, prototype refinement, and early governance planning.

The key contributions of this research are:
- **Framework Innovation (Core Contribution):** Introduces the PALEI framework as a structured method for integrating AI literacy, participatory engagement, fairness considerations, and explainability into urban and environmental resilience contexts. PALEI makes AI literacy and explainability designable, measurable, and governable in real community settings.
- **Applied Demonstration (Initial Implementation):** Establishes the Los Angeles wildfire risk assessment project as the first application of PALEI, using a virtual community town hall, focus-group–style discussion, scenario cards, and conceptual wireframes, to examine how explainability and trust are explored during early engagement.
- **Empirical Insight (Preliminary):** Provides early exploratory insight into how literacy-building, transparent communication, and localized explanation formats influence comprehension, fairness perception, and user confidence in AI-enabled tools, while broader data collection continues in subsequent phases.

The rest of the paper is structured around these contributions. The next section highlights the background scholarship that informs this work, situating the PALEI approach within existing research and identifying the gaps it seeks to address. The following section outlines the PALEI framework and its six stages, along with a description of how the approach will be applied in the case study communities. The paper then details the initial implementation of the first PALEI stages through the virtual community town hall and concludes with the contributions of this early engagement before model development begins.

## 1.0   RELATED WORK & THEORETICAL FOUNDATIONS

### 1.1 Human-Centered, Ethical, and Explainable AI
In human-centered design frameworks, such as ISO 9241-210:2019 and the Double Diamond model (Design Council 2019), iterative design and usability are emphasized, but AI-specific issues such as transparency, fairness auditing, uncertainty, and user literacy are not addressed. Ethical AI models, such as Shneiderman's Human-Centered AI (Shneiderman 2020) and Floridi et al.'s AI4People (Floridi et al. 2018), outline high-level principles but offer few concrete procedures for cultivating public trust and comprehension. Additionally, existing explainable AI tools, such as IBM's AIX360 (Arya et al. 2019) and Google's What-If Tool (Wexler et al. 2019), primarily support developer-facing model introspection rather than helping non-experts understand AI reasoning. Reviews of systematic literature confirm these gaps; as Abdul et al. (2020) highlight the lack of participatory practices in XAI, and Kaur et al. (2022) call for collaborative explanation design. Also, Weidinger et al. (2022) argue that technical transparency alone cannot address the broader social and ethical risks associated with AI systems. Table 1 summarizes the limitations of existing AI and design frameworks relevant to this study.

**Table 1**: Limitations of Existing Frameworks and Tools. Source: (Authors 2025)

| Framework / Tool | Core Focus | Limitation for This Study |
|---|---|---|
| ISO 9241-210; Double Diamond | Usability-focused HCD | No AI-specific interpretability, fairness auditing, or literacy process |
| HCAI; AI4People | Ethical principles | High-level; no procedures for evaluating trust or comprehension |
| Guidelines for Human–AI Interaction | Design heuristics | Limited mechanisms for participatory explanation testing |
| AIX360; What-If Tool | Developer-facing introspection | Not designed for public-facing understanding |
| Abdul et al., Kaur et al., Weidinger et al. | Reviews of XAI/trust | Identify need for participatory, governance-aware explainability |

### 1.2. Urban Fire Risk and the Need for Resident-Facing Interpretability
Urban fire risk emerges from dense development, aging infrastructure, and rapid expansion of the wildland–urban interface (WUI). In the United States, approximately 2 million acres enter the WUI each year and around 3,000 structures are destroyed each year (Goswami et al. 2024). In Los Angeles, development in fire-prone canyons elevates exposure and complicates evacuation ("Journal Article Explores Urban Resilience to Wildfires" 2025). Fire occurrence correlates with building density, access restrictions (Kim et al. 2024), and climate conditions such as heat, drought, and wind (Land Use Planning – Wildfire Risk to Communities 2025), with older, tightly spaced neighborhoods having disproportionately high fire incidences (Goswami et al. 2024). Communities in WUI regions are particularly vulnerable, since homes are typically ignited by neighboring burning structures rather than by vegetation itself (Mann et al. 2014; FEMA 2025). Thus, interpreting micro-level property characteristics in relation to neighborhood-scale transmission dynamics is important.

Urban wildfire vulnerability is consistently shaped by four dimensions across global contexts: *(1) urban design and access*, where dense blocks and compact street networks can accelerate spread and delay response, whereas well-planned routes and fire breaks can slow propagation (Zhou, Lin, and Wang 2025; Hao et al. 2023) *(2) land use and WUI patterns*, where residential areas and long-known fire-prone canyons expose millions of Californians to elevated risk (Hossain and Smirnov 2023); *(3) climate and weather*, including low humidity, extreme heat, and Santa Ana winds that intensify ignition and ember travel (Goswami et al. 2024; Yuan and Wylie 2024); and *(4) construction and materials*, where older buildings, degraded wiring, and flammable elements increase vulnerability and recommended protective measures remain unevenly enforced (Kim et al. 2024; Intini et al. 2020). So, because wildfire risk is highly localized, tools that can integrate both property-level conditions and neighborhood-scale patterns are needed.

### 1.3. AI for Fire Risk: Predictive Advances but Limited Public Interpretability
AI-based wildfire prediction has advanced quickly, with deep neural networks supporting 30-day forecasts (Zhou, Lin, and Wang 2025), ensemble–neural–spatial models improving performance in Chengdu (Hao et al. 2023), and SHAP-enhanced gradient boosting models aiding interpretability in Seoul (Kim et al. 2024). Remote-sensing systems such as VIIRS, MODIS, LightningCast, and digital twins further expand regional forecasting capacity (Huang, Wu, and Usmani 2022). Despite these advances, most systems are built for expert use. Public-facing platforms—FireCast, Wildfire Analyst, and the USFS Wildfire Risk to Communities portal—still require expert interpretation, lack parcel-level explanations, and rarely incorporate user-generated imagery (Conservation International 2023; Scott et al. 2020). Response apps such as FireAlerter and PulsePoint remain focused on post-ignition alerts rather than prevention (NFPA 2023; PulsePoint Foundation 2020). Overall, current tools do not help residents understand how micro-scale property conditions relate to neighborhood-level fire spread in ways that support action.

### 1.4. Policy and Governance: Structural Needs and Informational Gaps
Policy research highlights the need for coordinated governance among planners, fire services, and policymakers (FEMA 2025). Fire-risk assessments are recommended for zoning approvals and infrastructure planning, including evacuation capacity and water access (Calkin et al. 2023). AI-based analytics can support scenario evaluations, but implementation remains inconsistent. There are still gaps in hazard-informed zoning, compliance capacity, and mitigation enforcement that limit risk reduction. As a result, many governance strategies rely on residents' understanding and trust in the information they receive ("Designing Communities for Wildfire Resilience – SOM Research" 2025).

### 1.5. Synthesis: The Research Gap Addressed in This Paper

Across HCI/HCAI/XAI, urban fire science, AI prediction, and policy research, four persistent gaps emerge: **(1) from transparency to literacy**, as existing tools explain models to experts but do not help non-experts interpret or act on outputs (Arya et al. 2019; Kaur et al. 2022); **(2) from parcel to neighborhood spread**, since public-facing systems rarely connect household conditions to community-scale transmission pathways (Goswami et al. 2024; FEMA 2025); **(3) from accuracy to actionability**, because AI advances often lack resident-facing, uncertainty-aware explanation formats or prevention guidance (Lim et al. 2023; Yuan and Wylie 2024; Intini et al. 2020); and **(4) from principles to procedure**, as ethical AI frameworks offer high-level values but not participatory, measurable methods for evaluating comprehension, fairness, and trust (ISO 2019; Shneiderman 2020; Floridi et al. 2018). This gap motivated the PALEI framework, which emphasizes participatory, literacy-first, and explainability-centered evaluation prior to and during the early stages of predictive model or prototype tool development.

### 2.0 METHODOLOGICAL FRAMEWORK: PALEI

The Participatory AI Literacy and Explainability Integration (PALEI) framework forms the core methodological foundation of this research and guides the design, evaluation, and interpretation of the early-stage wildfire risk assessment tool for Los Angeles communities. A key goal of PALEI is to bridge persistent gaps between human–AI interaction and user-centered design, where usability often takes precedence over trust, fairness, and algorithmic understanding. In contrast, PALEI focuses on literacy, interpretability, and participatory governance as essential elements of responsible AI integration. Unlike conventional approaches that begin with a functional model, PALEI begins with a process of determining how people learn about, evaluate, and question emerging artificial intelligence systems before they are even implemented. This framework offers a variety of tools to support transparent, accountable, and community-centered design, including a multi-layered AI System Flow Map, Scenario Cards, AI Literacy Guides, AI Value Map, Explanation Personas, Explainability Prototype Tools, Bias and Fairness Monitoring Logs, and Governance and Explainability Tools. Figure 2 illustrates PALEI's six iterative stages.

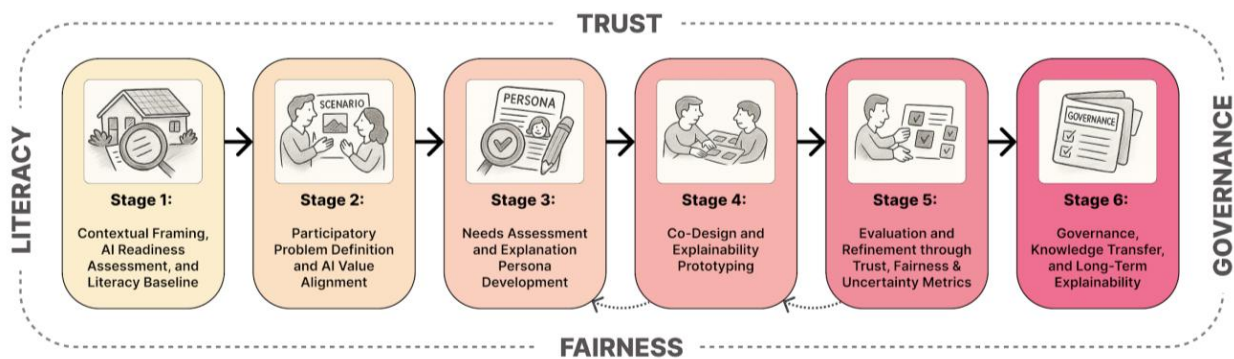

**Figure 2**: The six-stage PALEI framework outlining literacy building, value alignment, explanation-persona development, prototyping, evaluation, and governance planning for community-centered AI design. Source: (Authors 2025)

- **Stage 1: Contextual Framing, AI Readiness Assessment, and Literacy Baseline:**
  Stage 1 establishes a shared understanding of the system's purpose, constraints, and ethical considerations. As part of a *"Readiness and Fairness Audit"*, we assess data quality, representativeness, and potential bias sources. Based on these insights, the *"AI System Flow Map (Feasibility Layer)"*, is constructed to outline potential inputs, conceptual outputs, and uncertainty boundaries in an easily understandable manner. In this concise AI Literacy Primer, patterns, confidence, and uncertainty are described in plain language, so residents and stakeholders can participate meaningfully from the start.
- **Stage 2: Participatory Problem Definition and AI Value Alignment with Community:**
  In stage 2, community members and stakeholders engage in a structured dialogue about how AI should operate in their local context. Participants interact with the *"AI System Flow Map (Demonstration Layer)"*, *"Scenario Cards"*, and *"Literacy Guides"* which illustrate hypothetical AI outputs. Through facilitated discussion, they uncover expectations regarding fairness, preferred explanation styles, privacy, accountability, and the appropriate balance between human and machine judgment. Based on these inputs, an *"AI Value Map"* is created that provides the ethical and interpretive benchmarks for design decisions in the future.
- **Stage 3: Needs Assessment and Explanation Persona Development:**
  Stage 3 examines how different users make sense of conceptual AI explanations. Through surveys, interviews, and interactive work with the Scenario Cards, participants compare visual, numerical, and narrative explanation formats and reveal where uncertainty is clear or confusing. These insights are then developed into *"Explanation Personas"* that summarize different users' interpretive preferences, trust cues, fairness concerns, and comfort with ambiguity. The personas directly guide how future AI guidance should be communicated.

- **Stage 4: Co-Design and Explainability Prototyping:**
  In stage 4, AI-generated outputs are communicated in concrete ways. Using a "*Explainability Prototype Toolkit*", participants combine modular elements, such as confidence indicators, reasoning steps, and scenario-based examples, to compare alternative explanation formats. As a result of these co-design sessions, users can identify which approaches feel clear or confusing and how they can be supported in practice. Refinements are based on the *"Explanation Personas"* from Stage 3, ensuring they reflect actual interpretation needs. For more equitable revisions, a "*Bias and Fairness Monitoring Log*" captures points of misunderstanding and unequal interpretation.
- **Stage 5: Evaluation and Refinement through Trust, Fairness & Uncertainty Metrics:**
  Stage 5 shifts from designing explanations to assessing their real impact in subsequent phases of the PALEI process. Participants complete structured tasks that test whether they truly understand AI outputs, how they interpret uncertainty, and how they perceive fairness and trust. The measures will be combined with brief interviews. Additionally, stakeholders assess whether user interpretations align with operational and ethical expectations. The goal is to verify that the explanations are not only useful, but also credible and meaningful.
- **Stage 6: Governance, Knowledge Transfer, and Long-Term Explainability:**
  In Stage 6, the framework will be consolidated into a "*Governance and Explainability Toolkit*" containing updated literacy materials, improved explainability components, a "*Bias and Coverage Checklist*", and a "*Governance Handoff Plan*" for AI System Flow Maps. In order to prevent the AI tool from becoming a "black box" once deployed, these resources support ongoing transparency, staff training, and community feedback processes.

**PALEI in This Study**
In this research, PALEI is applied before any wildfire-risk assessment model is built, focusing on how residents understand conceptual AI explanations. This framework guides the development of all community engagement materials-literacy guides, scenario cards, explanation mockups, and survey instruments, that were used during the initial wildfire awareness event. This section describes how each PALEI stage has been and will be operationalized throughout the case study areas to inform the eventual development of a resident-facing wildfire risk assessment tool.

## 3.0 CASE STUDY APPLICATION: LOS ANGELES COUNTY

### 3.1. Conceptual Overview of the Proposed Tool
The wildfire risk assessment tool is envisioned as a resident-facing mobile application that enables homeowners to identify exterior fire vulnerabilities around their properties. In a future deployment, the app would allow users to scan features such as vegetation type and spacing, fuel proximity, gaps between homes, and siding or fencing materials. A vision–language model would then generate an intuitive risk score (e.g., 0–100), a brief explanation of why the score was assigned, an uncertainty range, and simple mitigation suggestions. A mature version of the system would also incorporate geospatial conditions—slope, wind exposure, fire history, and fuel continuity to show how property-level features contribute to neighborhood-scale vulnerability. For this study, all outputs are simulated rather than model-driven. The emphasis is on how risk information, uncertainty, privacy details, and mitigation guidance should be communicated so that the eventual tool is transparent, trustworthy, and aligned with community expectations. Table 2 outlines the planned activities associated with each PALEI stage.

### 3.2. Case Study Activities
The case study applies this conceptual tool to three high-risk communities, Altadena, Pasadena, and Pacific Palisades, to examine how residents and local officials interpret early forms of wildfire-risk information. Based on the PALEI framework, the initial fieldwork focuses on how participants engage with simulated AI outputs, including scenario cards, printed guides, wireframes, uncertainty examples, and mitigation suggestions. There is no predictive model available yet, so these materials act as stand-ins for evaluating clarity, trustworthiness, fairness, and perceived usefulness. This "human-and-AI-in-the-loop from the beginning" approach ensures that the future tool is shaped by community needs and lived experiences rather than by technical assumptions alone. Insights gained from this early engagement will inform subsequent stages of explanation-persona development, co-design, prototype refinement, and early governance planning. While the early engagement focuses on participants from Altadena, Pasadena, and Pacific Palisades, the broader case study targets residents across Los Angeles County through continued survey-based outreach in subsequent phases.

### 3.3. Synthesis
A six-stage case study of Los Angeles County shows how wildfire-risk information can be co-designed, tested, and refined with communities before a predictive model can be developed. The initial implementation included a focus-group–style community engagement, followed by an ongoing survey-based outreach intended to broaden participation across Los Angeles County beyond the three focal neighborhoods. These combined engagements are designed to reflect local interpretation needs and align with community expectations for clarity, fairness, and transparency.

**Table 2**: Summary of Activities Across PALEI Stages in the Los Angeles County Case Study. Source: (Authors 2025)

| PALEI Stage | Planned Activities |
|---|---|
| *Stage 1 – Local Risk Context, AI Feasibility & Literacy Baseline* | Neighborhood hazard scans (vegetation, spacing, slope, fire history); AI readiness and fairness audit; development of a simplified AI System Flow Map (Feasibility Layer); distribution of the AI Literacy Guide to establish a shared baseline before engagement. |
| *Stage 2 – Early Engagement, Scenario Reactions & Value Alignment* | Hosting a community town hall event; Resident review of Scenario Cards showing simulated AI scores, "why" explanations, uncertainty ranges, and mitigation steps; Printed Guide explaining AI score bands, uncertainty, and privacy; brief post-event interviews; short survey on clarity, trust, fairness, privacy, and preferences; stakeholder discussions on the expected role of AI, autonomy, and safeguards; development of the AI Value Map. |
| *Stage 3 – Explanation Needs Assessment & Persona Development* | Guided walkthrough of low-fidelity wireframes showing conceptual AI outputs; comparison of numeric, color-band, and narrative AI explanation formats; interviews with residents and stakeholders on defensible-space inspections and uncertainty communication; development of Explanation Personas reflecting how different users interpret early AI explanations. |
| *Stage 4 – Co-Design & Explainability Prototyping* | Co-design sessions where participants recombine modular AI explanation components (confidence indicators, summary statements, scenario-based reasoning); annotation activities to flag clarity vs. confusion; testing simplified vs. detailed AI reasoning formats; documentation of interpretive disparities through a Fairness & Bias Monitoring Log to refine how AI explanations should be communicated. |
| *Stage 5 – Evaluation Using Trust, Fairness & Uncertainty Metrics* | Interpretation tasks using high-fidelity mockups of AI-generated explanations; measurement of comprehension, interpretive efficiency, uncertainty understanding, and perceived fairness and trust; post-task interviews to assess whether AI explanations feel meaningful, credible, and decision-supportive from both resident and stakeholder perspectives. |
| *Stage 6 – Governance, Knowledge Transfer & Future Deployment Planning* | Creation of the Governance & Explainability Toolkit summarizing AI explanation standards; updated AI Literacy and onboarding materials; bias and coverage checklists; templates for long-term transparency and uncertainty communication; guidance for maintaining fairness and interpretability as the future AI model is developed and deployed. |

## 4.0 EARLY IMPLEMENTATION OF PALEI: COMMUNITY ENGAGEMENT AND INITIAL INSIGHTS

The Los Angeles County case study began with an online community wildfire awareness event planned for residents of Altadena, Pasadena, and Pacific Palisades. During the 90-minute virtual town hall, participants engaged in a structured session combining short survey segments, facilitated discussion, and interactive review of the AI System Flow Map, scenario cards, and conceptual app screens to operationalize PALEI Stages 1–2 by establishing a shared literacy baseline and gathering early reactions to simulated AI explanations. The unified survey instrument was intentionally divided into brief sections and administered at multiple points to reduce cognitive load and encourage reflection. Supporting materials, including the "Why We're Meeting" framing, guides, scenario cards, and survey design, were created directly based on the PALEI framework. Figure 3 presents illustrative scenario cards used to communicate simulated wildfire-risk conditions. The structured survey instrument used during the event remains open and is also distributed alongside the recorded session to broaden participation beyond the initial focus group. Consistent with standard focus group practice, the session involved seven participants, enabling in-depth discussion and collective reflection rather than statistical generalization. All activities were conducted under an approved Institutional Review Board (IRB) protocol at UNC Charlotte.

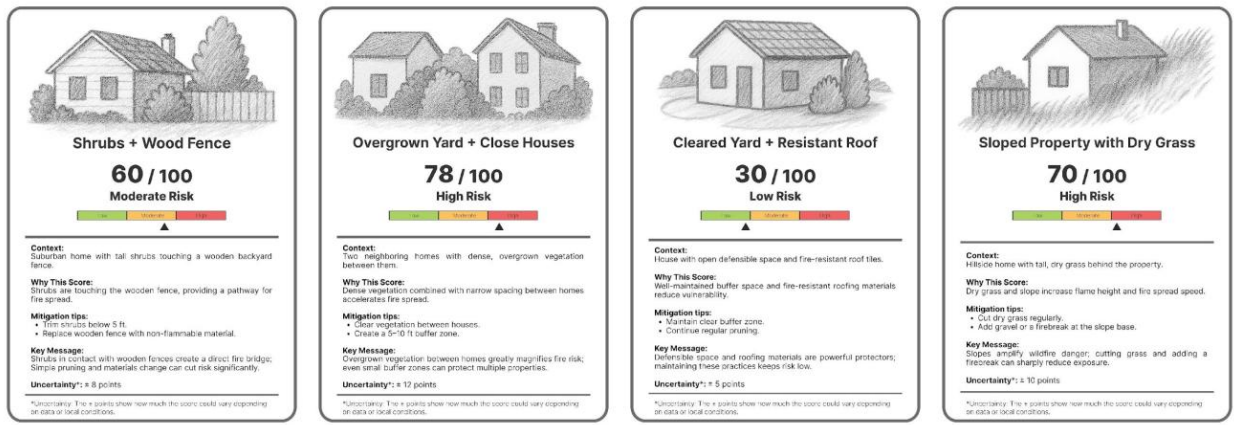

**Figure 3**: Scenario cards illustrating conceptual wildfire-risk conditions used during the initial engagement event. Source: (Authors 2025)

Because the predictive model has not yet been developed, the event first established a Stage 1 literacy and contextual baseline, introducing residents to core wildfire-risk concepts, the AI System Flow Map, and the role of micro-level conditions in neighborhood-scale fire spread. This shared foundation supported Stage 2 activities, where participants interpreted conceptual wildfire-risk information, including simulated risk scores, "why this score" explanations, uncertainty ranges, and mitigation suggestions. Using non-functional mid-fidelity wireframes and scenario-based examples, we introduced the conceptual tool and invited feedback on its clarity, trustworthiness, fairness, and privacy. Embedded survey segments collected structured responses on comprehension, uncertainty perception, fairness expectations, mitigation willingness, and preferred explanation formats. These responses will directly inform Stage 3 refinements and help identify which explanation approaches require adjustment.

The event followed a multi-phase format that combined brief presentations, facilitated group discussion, and embedded survey moments. This multi-phase structure allowed participants to first build a shared understanding of wildfire risk concepts, then react to visual assessment materials, and finally provide structured feedback on how risk information should be communicated and interpreted. As a result of this approach and putting community perspectives at the center of the early design process, the emerging tool is not only technically correct, but also socially relevant, practical, and aligned with the expectations of the public. In the long-term, we aim to strengthen wildfire preparedness through clearer communication, improved understanding, and shared resilience planning.

### 4.1. Early Contributions & Outcomes

The community wildfire awareness event served as an initial step in examining how residents interpret and respond to conceptual AI-generated wildfire-risk information. The town hall and accompanying survey were designed to explore the following early design dimensions:

- **Baseline risk perception:** Participants reported their current level of concern regarding wildfire risk in their neighborhood, providing contextual grounding for how the tool is interpreted and valued.
- **Interface usability and visual clarity:** Residents evaluated the overall look, layout, and flow of the conceptual interface to assess clarity, accessibility, and intuitive navigation.
- **Adoption potential:** Survey items measured the likelihood of using the app if available, offering early insight into perceived usefulness and behavioral intent.
- **Fairness and equity perception:** Participants assessed whether the tool would be useful and fair across different neighborhoods, informing equity considerations within model and interface design.
- **Governance and data-sharing willingness:** Questions examined willingness to share wildfire risk results and overall comfort with participation in community-level safety initiatives.
- **Privacy and trust concerns:** Respondents identified key concerns, including data security and result accuracy, that may influence confidence in system-generated assessments.

Preliminary survey responses (n = 28), including both live attendees and residents who viewed the recorded session and completed the same instrument, provide insight into how participants perceive and engage with the conceptual wildfire-risk interface. Most participants described themselves as slightly (46.4%) or moderately (28.6%) concerned about wildfire risk in their neighborhood, while 14.3% reported being very concerned. This distribution indicates that concern is present but generally moderate, aligning more with preparedness-oriented engagement than acute crisis response. Reactions to the interface were largely positive, with 24 respondents (85.7%) indicating that they "liked" or "strongly liked" the overall look and layout of the app. Adoption intent was similarly encouraging, as 71.4% reported that they would be likely or very likely to use the app if available. Perceptions of fairness were also positive, with 75.0% describing the tool as either "very fair" or "somewhat fair" across different neighborhoods.

Governance-related responses suggest a balance between willingness and concern. While 57.1% reported being willing or very willing to share their wildfire risk results to support community safety, 35.7% remained neutral and 7.1% expressed unwillingness. Privacy and data security emerged as the most frequently cited concern (57.1%), followed by concerns about accuracy and uncertainty (42.9%), underscoring the need for transparent communication and clear explanation mechanisms in future development. Although the sample is limited, the survey was conducted as part of a mixed-method user experience research process combining structured discussion and participatory evaluation. These exploratory findings provide practical direction for iterative co-design, explanation refinement, and governance planning within the PALEI framework. They support Stage 1–2 assumptions regarding a literacy-first engagement approach and indicate that privacy and uncertainty explanations should be designed early in the development process.

### CONCLUSION & FUTURE WORK

This paper presents a community-led framework for integrating artificial intelligence into wildfire risk assessment through a literacy-first, explainability-centered approach. By pairing the PALEI framework with an initial participatory case study in Los Angeles County, the work illustrates how communities can begin to shape the interpretive, ethical, and communicative foundations of an AI system long before any predictive model is created. The approach involves residents as co-designers whose lived experiences, concerns, and expectations inform the system from the earliest stages, rather than as passive recipients of technical outputs. The initial community engagement event, which included scenario cards, conceptual wireframes, literacy materials, and a structured survey, played a pivotal role in this process. Although no model currently exists, this initial engagement helped clarify how residents understand conceptual risk information, which explanation formats support trust and fairness, and what privacy considerations must be integrated into future design choices. In the next stages, explanation-persona development, early prototype design, and governance planning will be directed by these early insights. As the project progresses, PALEI will continue to guide iterative prototype testing, evaluation of explanation clarity and uncertainty comprehension, and development of an Explainability and Governance Toolkit. By focusing on public understanding, fairness, and transparency from the outset, the framework supports the development of AI systems that are not only technically credible, but also socially grounded, trustworthy, and aligned with the realities of the people and neighborhoods they are meant to protect. Looking ahead, the PALEI process is designed to be transferable to other climate-vulnerable regions facing similar challenges related to trust, interpretability, and uneven access to risk information. While Los Angeles County serves as the initial testbed, PALEI's literacy-first structure, explanation tools, and participatory evaluation components can be adapted to other environmental, infrastructure, or public-facing AI systems where community understanding and fairness are central to design and deployment.